\documentclass{aa}
\usepackage{natbib}
\bibpunct{(}{)}{;}{a}{}{,} 

\usepackage{graphicx}
\usepackage{txfonts}
\usepackage{upgreek}
\usepackage{comment}

\usepackage{tabularx} 
\usepackage{caption}
\usepackage{subfig}

\newcommand{\Myr}{\textrm{ Myr}}

\newcommand{\Kelvin}{\textrm{ K}}

\newcommand{\um}{\textrm{ } \upmu \textrm{m}}
\newcommand{\micron}{\textrm{ } \upmu \textrm{m}}

\newcommand{\COTWO}{$\textrm{CO}_2$}
\newcommand{\CO}{$\textrm{CO}$}

\newcommand{\HTWOO}{$\textrm{H}_2 \textrm{O}$}

\newcommand{\CHFOUR}{$\textrm{CH}_4$}
\newcommand{\NHTHREE}{$\textrm{NH}_3$}

\newcommand{\kms}{\textrm{ km } \textrm{s}^{-1}}

\newcommand{\COSIG}{$14.5$}
\newcommand{\WATERSIG}{$17.0$}
\newcommand{\BTSIG}{$25.0$}
\newcommand{\CONTRAST}{$2.5\times10^{-5}$}
\newcommand{\rawcontrast}{$1:240$}
\newcommand{\contrastratio}{40}

\begin{document}

\title{Medium-resolution integral-field spectroscopy for high-contrast exoplanet imaging:}
\subtitle{Molecule maps of the $\beta$ Pictoris system with SINFONI\thanks{Based on observations made with ESO Telescopes at the La Silla Paranal Observatory under programme ID 093.C-0626.}}

\author{H.~J.~Hoeijmakers\inst{\ref{inst1},\ref{inst2},\ref{inst3}}\and H.~Schwarz\inst{\ref{inst4}} \and I.~A.~G.~Snellen\inst{\ref{inst1}} \and R.~J. de Kok\inst{\ref{inst1},\ref{inst5},\ref{inst6}} \and M. Bonnefoy\inst{\ref{inst7}} \and G. Chauvin \inst{\ref{inst7},\ref{inst8}} \and A.M. Lagrange\inst{\ref{inst7}} \and J.H. Girard\inst{\ref{inst7},\ref{inst9}}}

\institute{Leiden Observatory, Leiden University, 2333CA Leiden, The Netherlands \email{hoeijmakers@strw.leidenuniv.nl}\label{inst1} \and Observatoire de Gen\`eve, Chemin des Maillettes 51, 1290 Versoix, Switzerland\label{inst2} \and Universit\"at Bern, Center for space and habitability, Gesellschaftstrasse 6, 3012 Bern, Switzerland\label{inst3} \and  Department of Astronomy and Astrophysics, 1156 High St., University of California, Santa Cruz, CA 95064, USA\label{inst4} \and SRON Netherlands Institute for Space Research, Sorbonnelaan 2, 3584 CA Utrecht, The Netherlands\label{inst5} \and Department of Physical Geography, Utrecht University, P.O. Box 80.115, 3508 TC Utrecht\label{inst6}, The Netherlands \and Universit\'e Grenoble Alpes, CNRS, IPAG, F-38000 Grenoble, France\label{inst7} \and Unidad Mixta Internacional Franco-Chilena de Astronm\'{i}a, CNRS/INSU UMI 3386 and Departamento de Astronom\'{i}a, Universidad de Chile, Casilla 36-D, Santiago, Chile\label{inst8} \and Space Telescope Science Institute, 3700 San Martin Drive, Baltimore, MD 21218, USA\label{inst9}}

\date{Received 26 - 02 - 2018 / Accepted XXXX }

\abstract {Angular Differential Imaging (ADI) and Spectral Differential Imaging (SDI) are well-established high-contrast imaging techniques, but their application is challenging for companions at small angular separations from their host stars.} 
{The aim of this paper is to investigate to what extent adaptive-optics assisted, medium-resolution ($R\sim5000$) integral field spectrographs (IFS) can be used to directly detect the absorption of molecular species in the spectra of planets and substellar companions when these are not present in the spectrum of the star.} 
{We analysed archival data of the $\beta$ Pictoris system taken with the SINFONI integral field spectrograph located at ESO's Very Large Telescope, originally taken to image $\beta$ Pictoris b using ADI techniques. At each spatial position in the field, a scaled instance of the stellar spectrum is subtracted from the data after which the residuals are cross-correlated with model spectra. The cross-correlation co-adds the individual absorption lines of the planet emission spectrum constructively, while this is not the case for (residual) telluric and stellar features.} 
{Cross-correlation with CO and \HTWOO ~models results in significant detections of $\beta$ Pictoris b with signal-to-noise ratios of \COSIG ~and \WATERSIG ~respectively. Correlation with a $T=1700$K BT-Settl model provides a detection with an SNR of \BTSIG. This in contrast to application of ADI, which barely reveals the planet. While the adaptive optics system only achieved modest Strehl ratios of 19-27\% leading to a raw contrast of \rawcontrast ~at the planet position, cross-correlation achieves a 3$\sigma$ contrast limit of \CONTRAST ~in this 2.5 hr data set, a factor $\sim40$ ~below the raw noise level at an angular distance of $0.36"$ from the star.} 
{Adaptive-optics assisted, medium-resolution IFS, such as SINFONI on the VLT and OSIRIS on the Keck Telescope, can be used for high-contrast imaging utilizing cross-correlation techniques for planets that are close to their star and embedded in speckle noise. We refer to this method as "molecule mapping", and advocate its application to observations with future medium resolution instruments, in particular ERIS on the VLT, HARMONI on the ELT and NIRSpec and MIRI on the JWST.}


\maketitle

\section{Introduction}

To directly image an extrasolar planet, the light of its host star must generally be suppressed by orders of magnitude. The techniques that have been developed to do this rely on a combination of precise wave front control to restore the diffraction limit of the telescope (adaptive optics) and coronographic techniques to attenuate the stellar light (phase mask, phase/amplitude pupil apodization combined with focal plane mask or interferometers).  In such imaging data, residual starlight is present in the form of speckles which may mimic point-source objects, confusing the detection of companions and planets.

Such residuals can be suppressed by adopting differential imaging strategies that assume that the residual pattern scales with wavelength (spectral differential imaging, SDI), with the polarimetric state (polarimetric differential imaging, PDI) or is stable in time (angular differential imaging, ADI). Subsequent post-processing algorithms (cADI, LOCI, PCA, ANDROMEDA) aim to optimize the residual attenuation while conserving the planatery signal \citep[see e.g.][for reviews]{Guyon2011,Mawet2012,Chauvin2016}.

So far, direct imaging detections are generally limited to a specific part of the exoplanet population: Young gas giants in wide orbits that glow by radiating out the internal heat that remained from the time of their formation. These planets can be resolved from their host stars owing to their relatively large mutual angular separation and brightness at infrared wavelengths. Such planets can generally only be observed during the first few tens of millions of years of their lifetime, after which they have cooled too much to be detected with current facilities \citep[see e.g.][for a review]{Bowler2016}.

The desire to image cooler (i.e. older or less massive) planets that are closer to their host star is fuelling the development of new instruments, as well as new observing and data analysis techniques. The arsenal of high-contrast imaging facilities has recently been expanded by the deployment of dedicated planet finding instruments such as the Gemini Planet Finder \citep{Macintosh2006} at the Gemini Telescope, SPHERE \citep{Beuzit2008} on the European Very Large Telescope (VLT), and SCExAO \citep{Jovanovic2015} on the Subaru telescope. These instruments also have low-resolution (R$\sim30-400$) spectroscopic capability - allowing the spectal characterization of directly imaged planets and achieving planet-to-star contrast ratios down to $10^{-6}$ at $0.2"$ angular separation \citep[see e.g.][]{Macintosh2014,Ruffio2017,Mesa2017,Currie2017}.

Multiple studies indicate that when high-contrast imaging is combined with high-dispersion spectroscopic techniques, the achieved contrasts can be significantly enhanced \citep[see e.g.][]{Sparks2002, RiaudSchneider2007, Kawahara2014,Snellen2015,Luger2017,Wang2017,Lovis2017}. This strategy assumes that the faint planet and the much brighter star have spectral properties that are distinctly different at high spectral resolution; notably because of molecular absorption bands in the spectrum of the planet. Because the planet is close to the star, its spectrum is deeply embedded in speckle-noise, but it can be extracted using cross-correlation: The cross-correlation co-adds the individual absorption lines of the planet constructively but not stellar and telluric features, or at different radial velocities. This method has been applied successfully for the first time at high spectral resolution, albeit using the CRIRES slit spectrograph that probes only one spatial dimension. This resulted in the measurement of the spin rate of $\beta$ Pictoris b \citep{Snellen2014}.

In this paper we investigate to what extent this method can be applied to observations from adaptive-optics assisted, medium-resolution ($R\sim5000$) integral field spectrographs (IFS) - targeting molecular species in a planet atmosphere that are not present in the star, throughout the two-dimensional field of view. The use of cross-correlation techniques on IFS data was first performed by \citet{Konopacky2013} and \citet{Barman2015} to detect \HTWOO, \CO ~and \CHFOUR ~in the atmospheres of HR 8799 b\&c using the OSIRIS integral field spectrograph at the Keck Observatory. At $1.7"$ and $0.96"$, HR 8799 b\&c are widely separated from their host star \citep{Marois2008}. Such angular distances are well resolvable by modern adaptive-optics systems, and the speckle-pattern can be effectively removed using ADI or SDI-based methods. However, the application of ADI and SDI are less effective at smaller angular separations because the effects of field rotation and wavelength-dependencies are limited \citep{Fitzgerald2006,Lafreniere2007,Marois2008, Rameau2015}.

We apply the cross-correlation technique to archival K-band SINFONI image cubes of the $\beta$ Pictoris system, and are able to extract the spectral signature of the planet $\beta$ Pictoris b while effectively removing the diffraction pattern of the star in which the planet is embedded. We refer to this technique as \textit{molecule mapping}, because it produces two-dimensional cross-correlation maps that indicate the presence of certain molecular signatures for each location in the two-dimensional image.

In Section 2 we summarize the $\beta$ Pictoris system. Section \ref{sec:observations} describes the archival SINFONI observations used in this analysis and the data reduction. The cross-correlation procedure is described in Section \ref{sec:analysis}, followed by the resulting cross-correlation images in Section \ref{sec:results}, including a comparison with ADI using classical ADI \citep{Marois2006} and the LOCI algorithm \citep{Lafreniere2007}. Paragraphs \ref{sec:JWST} and \ref{sec:HARMONI} highlight the potential application of molecule mapping with upcoming medium-resolution integral-field instruments on the VLT, ELT and JWST, after which the paper is summarized and concluded in Section \ref{sec:conclusion}.

\begin{table}[h!]
\begin{tabularx}{\linewidth}{lll}
& \textbf{Symbol} & \textbf{Value}\\ 
\hline
Star:\\
Visible magnitude$^a$					& $V$ 					& 3.86\\ 
K-band magnitude$^a$					& $K$					& 3.48\\
Distance $\textrm{(pc)}^b$ 				& $d$ 					& $19.44 \pm0.05$\\
Effective temperature $(\textrm{K})^c$ 	& $T_{\textrm{eff}}$ 	& $8052 \pm30$\\
Mass $(M_{\odot})^d$	   				& $M_*$ 				& $1.85^{+0.03}_{-0.04}$\\
Metallicity $(\textrm{dex})^c$			& $\left[\textrm{M}/ \textrm{H} \right]$  & $0.05 \pm0.06$\\
Age $\textrm{(Myr)}^e$					& 						& $24 \pm3$\\ 
Rotation velocity $\textrm{(}\kms\textrm{)}^f$ & $v\sin i$      & 130 \\
Systemic velocity $ \textrm{(}\kms\textrm{)}^g$ 	& $v$      & $20.0\pm0.7$ \\

\hline
Planet:\\
Luminosity $\left(\log\frac{L}{L_{\odot}}\right)^h$ & $L_p$				&  $-3.78 \pm 0.03$\\
K-band contrast$^i$					&  $\Delta K$			  &    $9.2 \pm 0.1$\\
Effective temperature $(K)^h$							& $T_{\textrm{eff}}$&  $1724 \pm 15$\\
Surface gravity $(\log g)^h$					&					&  $4.18 \pm 0.01$\\
Angular separation $\textrm{(mas)}^i$			&					&  $356.5 \pm 0.9$\\
Position angle	$\textrm{(deg)}^j$ 	 			&					&  $213.0 \pm 0.2$\\
\end{tabularx}
\caption[lalaaa]{Summary of the properties of the star $\beta$ Pictoris (upper part) and its planet (lower part).\\
$a$: \citet{Ducati2002}.\\
$b$: \citet{VanLeeuwen2007}.\\
$c$: \citet{Gray2006}.\\
$d$: \citet{Wang2016}. The stellar mass was inferred from their fit to the total mass of the system minus the mass of the planet. Although the quoted confidence intervals correspond to $1\sigma$ uncertainties, the posterior distributions are non-Gaussian. \\
$e$: \citet{Bell2015}.\\
$f$: \citet{Royer2007}.\\
$g$: \citet{Gontcharov2006}.\\
$h$: \citet{Chilcote2017}. These authors measured the luminosity from their GPI datasets, but inferred the other parameters using a hot-start evolutionary model by \citet{Baraffe2003}. The reported statistical errors do not account for the model-dependency of these values.\\
$i$: \citet{Bonnefoy2011}.\\
$j$: Measurements obtained by \citet{Wang2016} on November 8, 2014.}
\label{tab:properties}
\end{table}

\section{The $\beta$ Pictoris system}
In 1984, $\beta$ Pictoris was the first star to be found to host a debris disk and which was associated with planet formation processes \citep{Smith1984}. The warped structure of the disk led to the inference of a possibly planetary mass companion \citep{Scholl1993,Roques1994,Lazzaro1994,Burrows1995,Lecavelier1996,Mouillet1997,Augereau2001}, which was discovered in 2008  via direct imaging \citep{Lagrange2009,Lagrange2010}. The system is a member of the nearby $\beta$ Pictoris moving group which has an estimated age of $24 \pm3\Myr$ \citep{Bell2015}. The bolometric luminosity and effective temperature of the planet were recently estimated by \citet{Chilcote2017} at $\log\left(\frac{L_p}{L_{\odot}}\right) = -3.78 \pm 0.03$ and $1724 \pm 15 K$ respectively. From these, the radius and mass of the planet were inferred to be $12.9 \pm 0.2 M_{\textrm{J}}$ and $1.46\pm0.01 R_{\text{J}}$ respectively \citep{Chilcote2017}, assuming a hot-start evolution model \citep{Baraffe2003}. These properties are summarized in Table \ref{tab:properties}.

$\beta$ Pic b moves on a 20--26-year orbit that is highly inclined with respect to the line of sight from Earth \citep{Wang2016}. This high inclination has recently triggered a renewed interest in this system because the Hill sphere of the planet was predicted to transit the star between April 2017 and January 2018 \citep{Lecavelier2016,Wang2016,Kenworthy2017}. \citet{Snellen2014} measured the projected equatorial rotation velocity of the planet, inferring a length of day of $\sim$8 hours.

\begin{table}[h!]
\begin{tabularx}{\linewidth}{lrr}
 Observing night & Sept. 10, 2014 & Sept. 11, 2014\\ 
\hline
$N_{\textrm{exp}}$ & 24 & 30 \\
DIT (s) & 60 & 2 \\
NDIT & 4 & 50 \\
Airmass & 1.34 - 1.15 & 1.49 - 1.14 \\
PA (deg) & 340.7 - 304.6 & 355.7 - 304.3 \\
Strehl ratio (\%) & 19 - 27 & 20 - 25 \\
\end{tabularx}
\caption{Overview of the observations of the the $\beta$ Pictoris system of September 10 and 11, 2014, showing the number of exposures $N_{\textrm{exp}}$, the exposure time per dithering position (DIT), the number of dithering positions combined to make each science frame (NDIT), and the range of airmasses, position angles on sky and approximate Strehl ratios achieved.}
\label{tab:observations}
\end{table}

\section{Observations}\label{sec:observations}
\begin{figure*}
   \centering
   \includegraphics[width=0.8\linewidth]{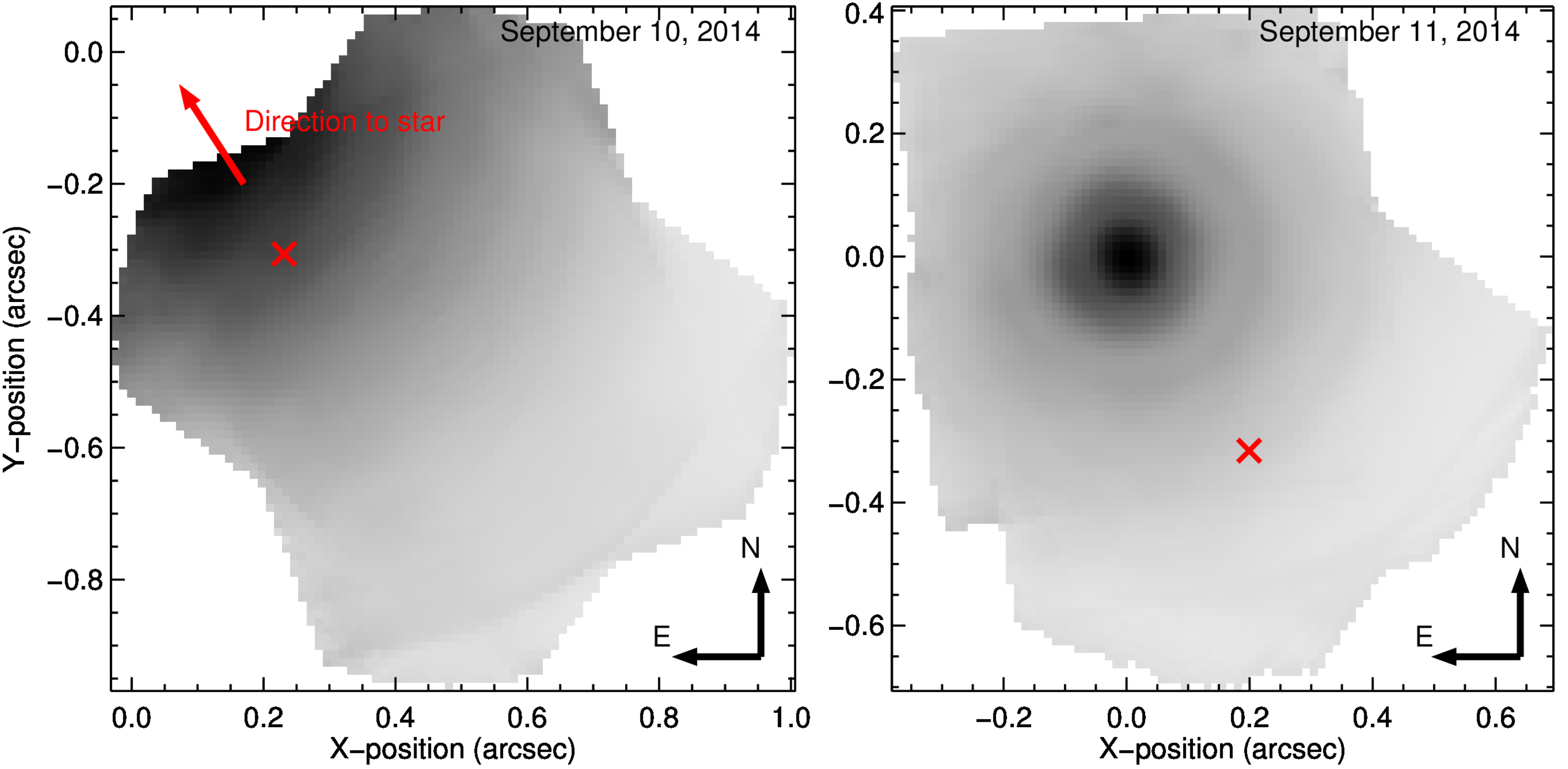}
   \caption{Wavelength-averaged images of the $\beta$ Pictoris SINFONI data, obtained by de-rotating and stacking the exposures of each night, and taking the median flux of the spectrum at each image location. The expected position of $\beta$ Pictoris b is indicated by the red crosses. The coordinate system is relative to the position of the star.}%
\label{fig:whiteframes}
\end{figure*}

The $\beta$ Pictoris system was observed with the SINFONI IFS \citep{Eisenhauer2003,Bonnet2004} mounted on ESO's Very Large Telescope in K-band on the nights of September 10 and 11, 2014 as part of ESO programme 093.C-0626(A) (P.I.: Chauvin). The observations on both nights were taken in pupil tracking mode which allows the field to rotate during the observing sequence, facilitating Angular Differential Imaging. The 32x64 pixel images were taken in the highest spatial resolution mode, with each pixel covering 0.0125" by 0.025", providing a field of view (FOV) of $0.8"$ by $0.8"$ on sky. At a resolving power of $R\sim5000$ the SINFONI spectra have a wavelength coverage from 1.929$\micron$ to 2.472$\micron$ with a sampling of $0.25$ nm. At the time of these observations the planet was separated from the central star by approximately $0.36"$ \citep{Wang2016}. On the first and second night, 24 and 30 science frames were obtained respectively, spanning a total duration of 2.5 hours (see Table \ref{tab:observations}). During both nights, the seeing varied between $0.7"$ to $0.9"$, and the MACAO AO system delivered modest Strehl ratios between $19\%$ and $27\%$ \citep{Bonnet2003}. The observations on the first night were preceded by a few acquisition images after the star was placed just outside the FOV for the rest of the sequence, enabling longer exposure times. During the second night both star and planet were observed continuously (see Fig. \ref{fig:whiteframes}).

\begin{figure*}
   \centering
   \includegraphics[width=0.8\linewidth]{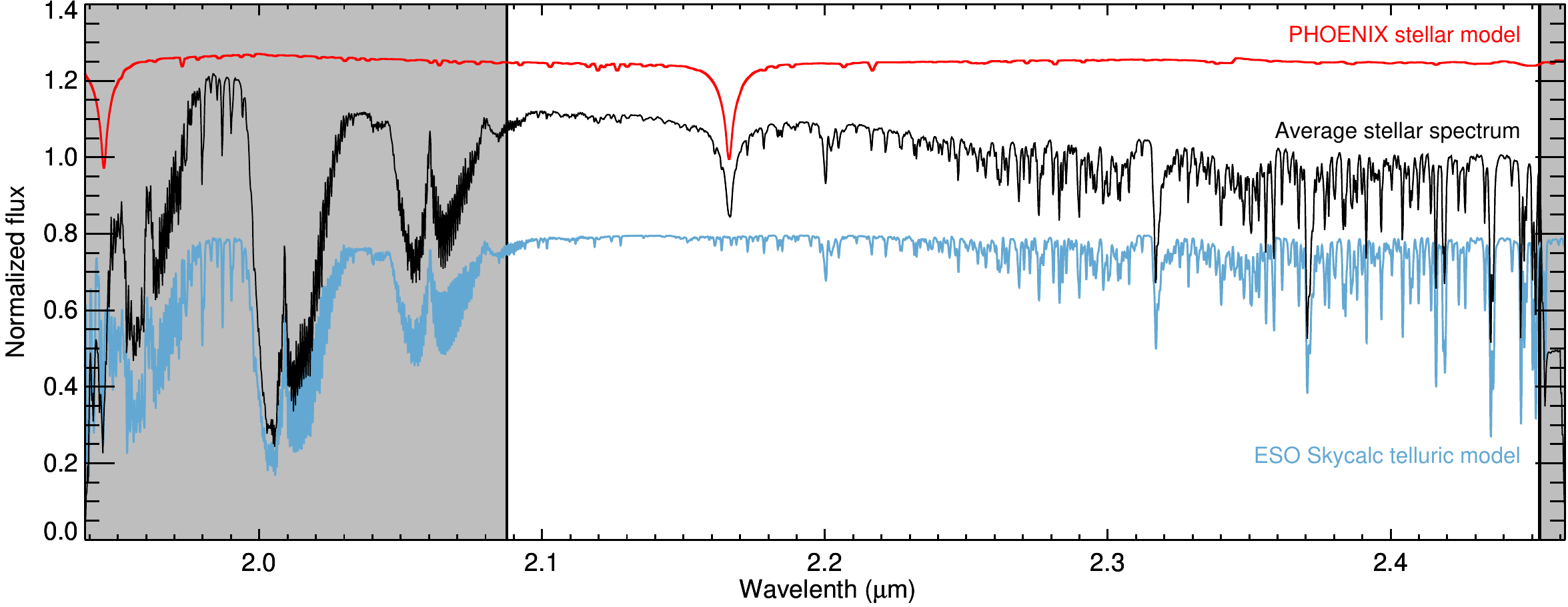}
   \caption{The average stellar spectrum of the master cube of the second night (black), compared to a continuum-normalized rotation-broadened PHOENIX model in red \citep{Husser2013} and a telluric transmission spectrum as obtained using ESO SkyCalc in blue \citep{Noll2012,Jones2013}, both convolved to the spectral resolution of SINFONI. There are few stellar absorption lines in this wavelength range, and the data is dominated by telluric absorption bands due to water, \COTWO and methane. The grey regions at the edges of the waveband were discarded due to strong telluric bands of \HTWOO ~and \COTWO ~or bad pixels that are close to the edges of the detector.}%
\label{fig:raw_data}
\end{figure*}

\begin{figure}
   \centering
   \includegraphics[width=0.9\linewidth]{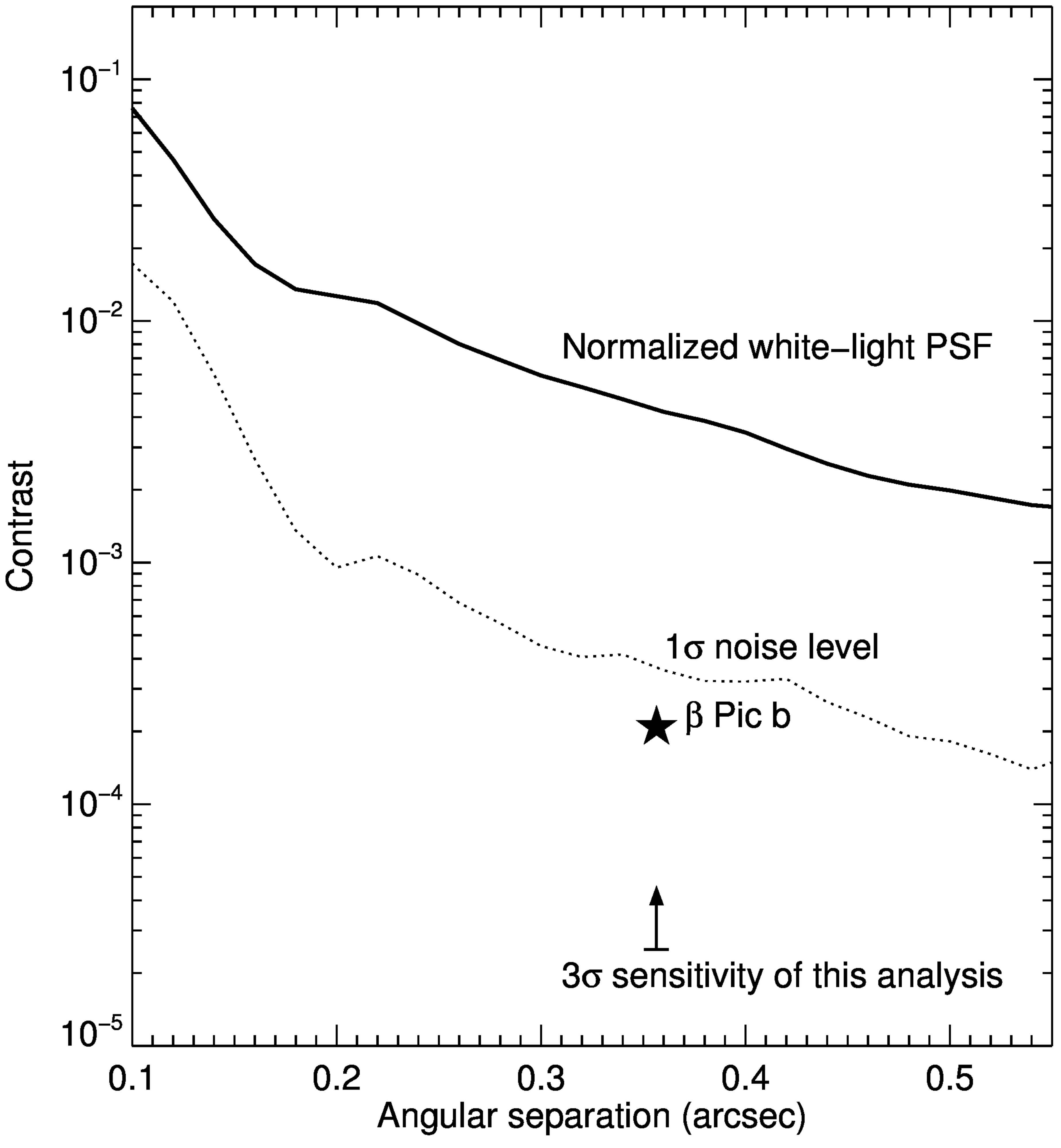}
   \caption{Raw-contrast curve obtained by averaging the flux from the second night in annuli with widths of 20mas around the central star, normalized by the peak flux. The standard deviation of the flux in these annuli is plotted with the dashed line, and is dominated by speckle noise. The $3\sigma$ sensitivity limit of the analysis presented in this work is derived from our SNR=\BTSIG ~detection of the planet, assuming that the actual K-band contrast of $\beta$ Pic b is $K=9.2$ \citep{Bonnefoy2011}.}%
\label{fig:contrast_curve}
\end{figure}

The raw data were downloaded from the ESO Science Archive Facility and reduced using version 3.0.0 of the SINFONI pipeline. The reduction pipeline produces a three-dimensional image cube for each science observation, with sky position in the $x$ and $y$ directions and wavelength in the $z$ direction. The pipeline reduced data cubes contain NaN values at the edges of the waveband and at the location of known bad pixels. To reject these areas, we only consider wavelengths between 2.088$\micron$ and 2.452$\micron$ and flag any remaining NaN values (see Fig. \ref{fig:raw_data}). From each science-exposure we obtain the wavelength-averaged two-dimensional image. The spatial location of the star was found by fitting a two-dimensional Gaussian profile to a region of $10 \times 10$ pixels around the maximum of the PSF. We use this location as a pivot to de-rotate all image-cubes by their respective position angle and to align them to a common frame, both using linear interpolation. We co-added all image cubes of each night to obtain two master image-cubes, excluding all regions within a range of 5 pixels ($\sim$60 mas) from the edges of the individual frames. The wavelength-averaged images of these two master cubes are shown in Fig. \ref{fig:whiteframes}. From the wavelength-averaged image of the second night we measured the raw PSF contrast and the noise as a function of angular separation by measuring the mean and the standard deviation of the flux in annuli centered on the star, normalized by the peak stellar flux (see Fig. \ref{fig:contrast_curve}). The contrast in intensity between the flux at the location of the planet with respect to the peak intensity of the PSF is measured to be 1:240, and the contrast between the $1\sigma$ noise level and the PSF peak is $3.6 \times 10^{-4}$.

We find a spurious radial velocity variation of the telluric absorption lines in the background spectra of both nights, increasing from $+20$ to $+30\kms$ during the observations, indicating instability in the wavelength solution at the 1 pixel level. We correct these anomalies by shifting each datacube to the rest-frame of the telluric spectrum.

\section{Data analysis}\label{sec:analysis}
\subsection{Removal of starlight}\label{sec:cleaning}
\begin{figure}
   \centering
   \includegraphics[width=\linewidth]{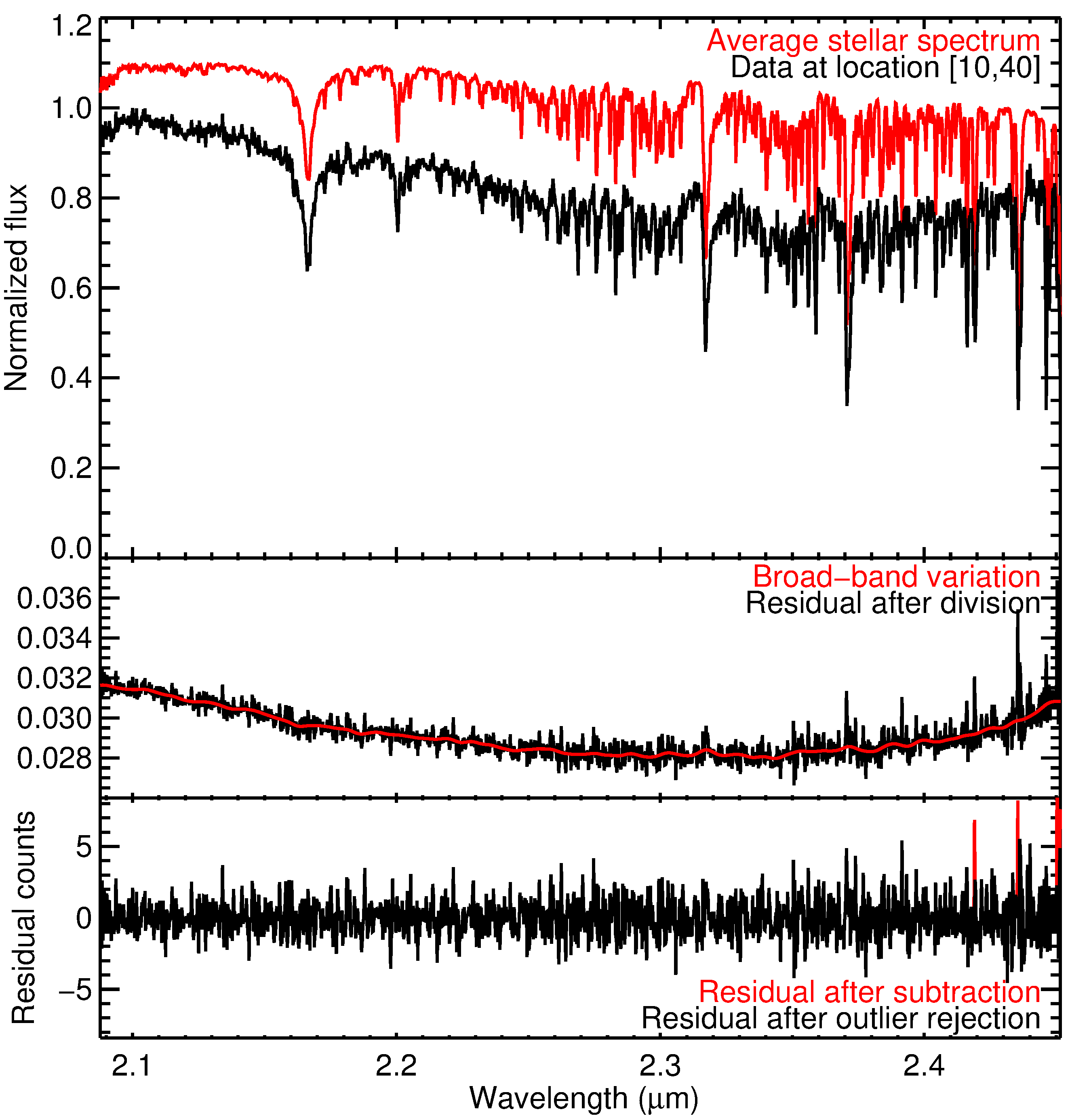}
   \caption{Stepwise removal of starlight from each spatial location in the data cubes. \textbf{Upper panel:} The master stellar spectrum (red) compared to the local observed spectrum (black) at spatial pixel location [10,40] of exposure \#7 of the second night. Both are normalized and offset to allow visual comparison.  
\textbf{Middle panel:} Ratio of the local to the master spectrum before (black) and after Gaussian smoothing (red). This smoothed residual is used as a proxy for the local wavelength-dependence of the stellar diffraction pattern. 
\textbf{Bottom panel:} Residuals obtained after subtracting the master stellar spectrum multiplied by the proxy from the local observed spectrum. Pixel values that are more than $6\sigma$ away from the mean are rejected (red). SYSREM is subsequently applied to these residuals (not shown).}%
\label{fig:cleaning}
\end{figure}

The flux at each spatial position in the data cube is dominated by the spectrum of the star, showing a few stellar but mainly telluric absorption features. 
The steps to remove this starlight are shown in Fig. \ref{fig:cleaning}. First, a high signal-to-noise master stellar spectrum is created by combining the spectra from the 1\% (20) brightest pixels by first normalizing them to the same flux level and then taking the mean of these spectra at each wavelength while rejecting  $>6\sigma$ outliers. 

For each spatial position in the data cube we subsequently divide the spectrum by this master and apply a Gaussian smoothing filtering with a 1$\sigma$ width of 10 wavelength steps ($\sim300\kms$). The resulting function is used as a proxy for the local wavelength-dependence of the stellar diffraction pattern. We multiply the stellar master spectrum with this proxy and subtract it from the local spectrum, rejecting any remaining $>6\sigma$ outliers. This effectively removes the starlight at each position and wavelength, as well as cosmic ray hits and bad pixels that were not corrected by the pipeline.

At this stage, residual structures remain in the image at the position of strong absorption lines, most importantly the strong telluric \CHFOUR ~feature at $2.32\um$ that dominates the absorption spectra of both hot and cold \CHFOUR ~gas. Such residuals are caused by mismatches between the local spectra and the stellar master spectrum. This effect notably occurs near the edges of individual frames and at the location of the star, where the signal-to-noise of the residuals is high. We attribute such systematic residuals to stray light in the instrument that was not corrected by the data reduction pipeline, diluting the spectra and hence decreasing the apparent depth of absorption features in the stellar/telluric spectrum, as well as remaining inaccuracies in the wavelength solution.

To mitigate these effects we unwrap the three-dimensional residual cube (i.e. the residuals after subtraction of the stellar spectrum) of each exposure to its original two-dimensional format and remove correlated noise patterns by iteravely applying the SYSREM algorithm \citep{Tamuz2005} on these two-dimensional frames. We found that applying more than 8 iterations did not improve the end result further. After application of SYSREM, the two-dimensional frames are folded back into three-dimensional cubes. Finally, these cubes are derotated and aligned as described in Section \ref{sec:observations} and co-added into two master residual data-cubes, one for each night, onto which we apply the cross-correlation analysis.

\subsection{Cross correlation with model templates}
We searched for the presence of \CO, \HTWOO, \NHTHREE ~and \CHFOUR ~at each spatial location in the data cube by cross-correlating each spectrum in the residual data-cubes with a model template of each of these molecules, corresponding to $\sim 1200\Kelvin$ emission models of the day side HD 189733 adopted from \citet{deKok2014}. In addition we cross-correlated with a grid of pre-computed BT-Settl model spectra\footnote{The BT-Settl model grid was obtained from https://phoenix.ens-lyon.fr/Grids/BT-Settl/CIFIST2011\_2015/FITS/} with temperatures $<3000\Kelvin$ and varying in surface gravity from $\log(g) = 2.5$ to 5.5 \citep{Allard2011}.

\begin{figure*}
   \centering
   \includegraphics[width=0.81\linewidth]{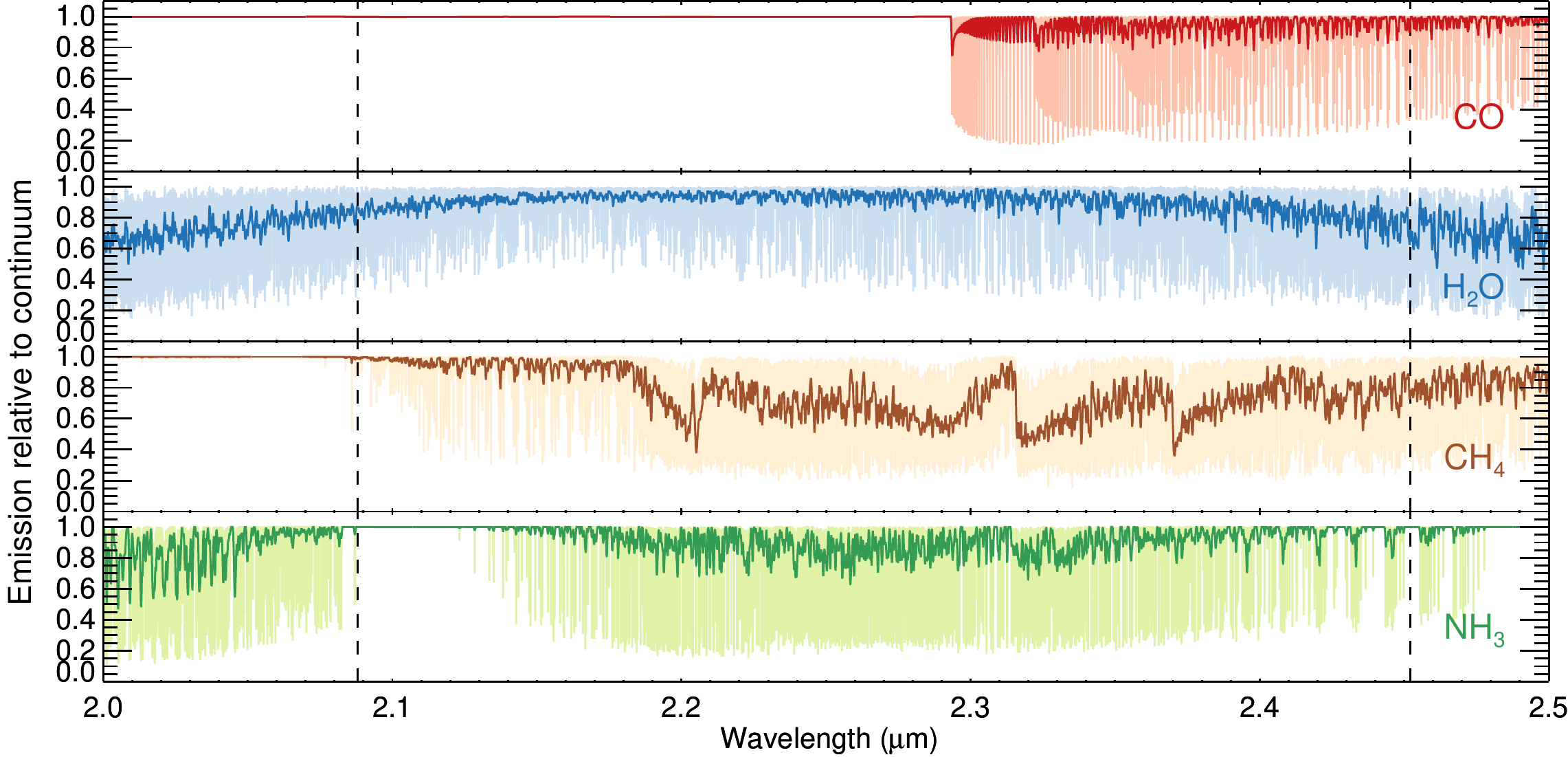}
   \caption{Model templates of \CO, \HTWOO, \CHFOUR ~and \NHTHREE ~at high ($R \sim 10^6$) spectral resolution (light colour) and convolved to a spectral resolution of $R=5000$ (dark colour). The vertical dashed lines indicate the wavelength range of the data.}%
\label{fig:templates}
\end{figure*}

Prior to cross-correlation we fit and subtract the continuum of these model spectra to eliminate low-frequency variations and convolve them to a spectral resolution of $R=5000$ (see Fig. \ref{fig:templates}). These models are then cross-correlated with each residual spectrum over a range of velocities of $v \pm 2500\kms$ in steps of $10\kms$. The cross-correlation of these two data cubes therefore results in two new  cubes that contain the cross-correlation coefficients of each spectrum for 501 steps in radial velocity. 

The closer the model template matches the real emission spectrum of the planet, the larger the cross-correlation function at the location and radial velocity of planet will be. At all other locations and radial velocities, no significant cross-correlation signals are expected. The systematic radial velocity of the $\beta$ Pictoris system is $+20\pm0.7\kms$ \citep{Gontcharov2006}. Using the orbital solution of \citet{Wang2016}, we calculate the instantaneous radial velocity of $\beta$ Pic b to be $\sim-9.5\kms$ with respect to the star. The barycentric velocity is $-7.9\kms$ so the cross-correlation signal of the planet is expected to occur at a radial velocity of $\sim3\kms$, i.e. at the central slice of the cross-correlation cubes.

After cross-correlation, the cross-correlation cubes associated with each night are aligned based on the known location of the star in the field (see Section \ref{sec:observations}) and subsequently averaged, weighed by the square root of the total exposure time in each night. This yields a single cross-correlation cube for each model template, the central slice of which contains the molecule maps of $\beta$ Pictoris b. 

\subsection{ADI Analysis}
We also analysed the data by means of angular differential imaging (ADI) to compare the performance of the molecule mapping technique. We applied the LOCI algorithm \citep{Lafreniere2007} on the sequence obtained during the second night. LOCI was applied independently at each wavelength with a separation criterion of 1 full-width-at-half-maximum, an optimization area of 300 FWHM and a geometry parameter of 0.5, after which the images at the individual wavelengths were stacked. During the first night of data the telescope was offset such that the star was located outside the field of view. This field assymetry causes the PSF of the star to evolve during the night due to field rotation and variable atmospheric refraction, hindering the application of LOCI. We therefore applied the classical ADI algorithm \citep{Marois2006} to the data obtained during the first night instead.

\begin{figure*}
   \centering
   \includegraphics[width=0.7\linewidth]{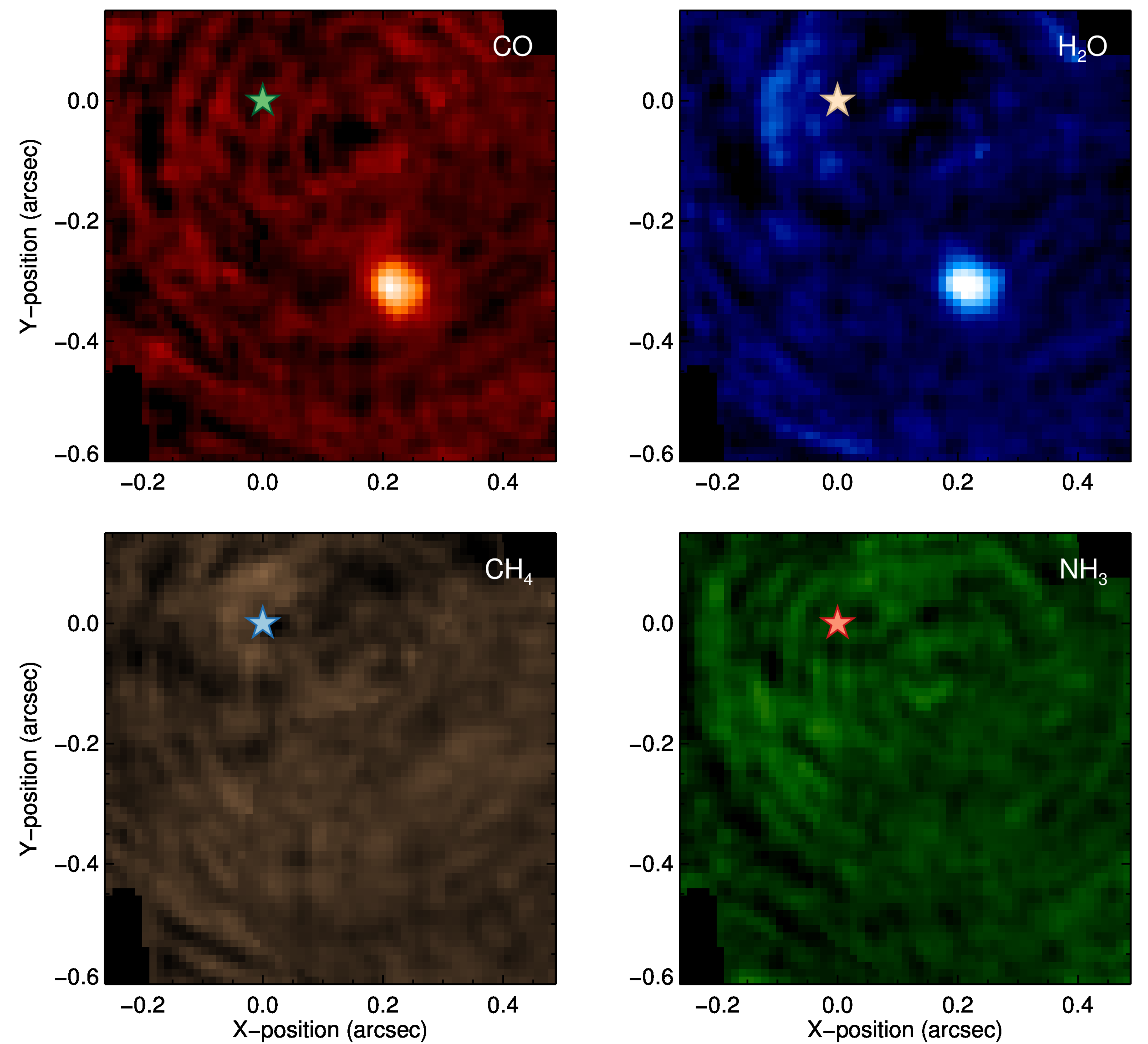}
   \caption{Molecule maps of \CO, \HTWOO,\CHFOUR ~and \NHTHREE ~at $v_{\textrm{sys}}=0\kms$. In all four panels, the colours scale linearly between cross-correlation values of -0.05 (black) to +0.2 (white). A cross-correlation enhancement caused by the planet is detected at a signal-to-noise ratios of \COSIG ~and \WATERSIG ~in the maps of \CO ~and \HTWOO ~respectively, but not in those of \CHFOUR ~and \NHTHREE.}%
\label{fig:xcor_maps}
\end{figure*}

\begin{figure*}
   \centering
   \includegraphics[width=0.9\linewidth]{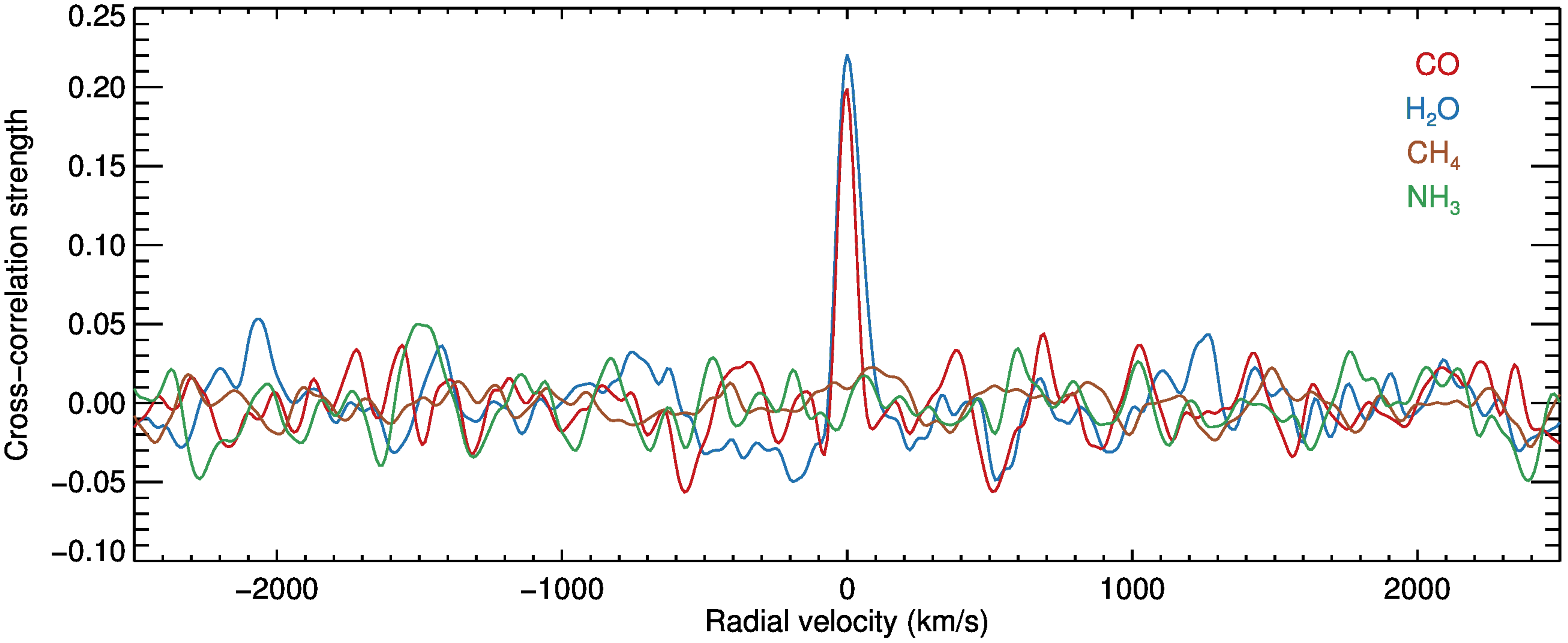}
   \caption{One-dimensional cross-correlation functions of \CO, \HTWOO, \CHFOUR ~and \NHTHREE ~at the location of the planet}.
\label{fig:xcor_plots}
\end{figure*}

\section{Results and Discussion}\label{sec:results}
\subsection{Molecule maps of the $\beta$ Pictoris system}
The molecule maps at the radial velocity of the planet are shown in Fig. \ref{fig:xcor_maps}. The maps of  \CO ~and \HTWOO ~show a significant signal at the expected position of the planet, while the planet is not detected in the maps of \NHTHREE ~and \CHFOUR. ~The one-dimensional cross-correlation functions at the position of the planet are shown in Fig. \ref{fig:xcor_plots}. To determine the signal-to-noise ratio of these signals we measure the average standard deviation of the cross-correlation function in an annulus between 8 to 11 pixels away from the peak signal, at radial velocities more than $\pm 250 \kms$ away from the planet. This is done to avoid systematic variations in the cross-correlation function at the location of the planet due to autocorrelation, which is not negligible due to the strength of the main cross-correlation peak. In this way we establish signal-to-noise of ratios for \CO ~and \HTWOO ~of \COSIG ~and \WATERSIG ~respectively. The non-detections of \CHFOUR ~and \NHTHREE ~are consistent with the effective temperature of the planet, which at 1724 K is expected to be too high for these molecules to be present at significant quantities unless the planet is rich in respectively carbon or nitrogen \citep{Burrows2006,Hubeny2007,Zahnle2014,Heng2016a,Heng2016b,Moses2016,Todorov2016}. 

\begin{figure*}
   \centering
   \includegraphics[width=\linewidth]{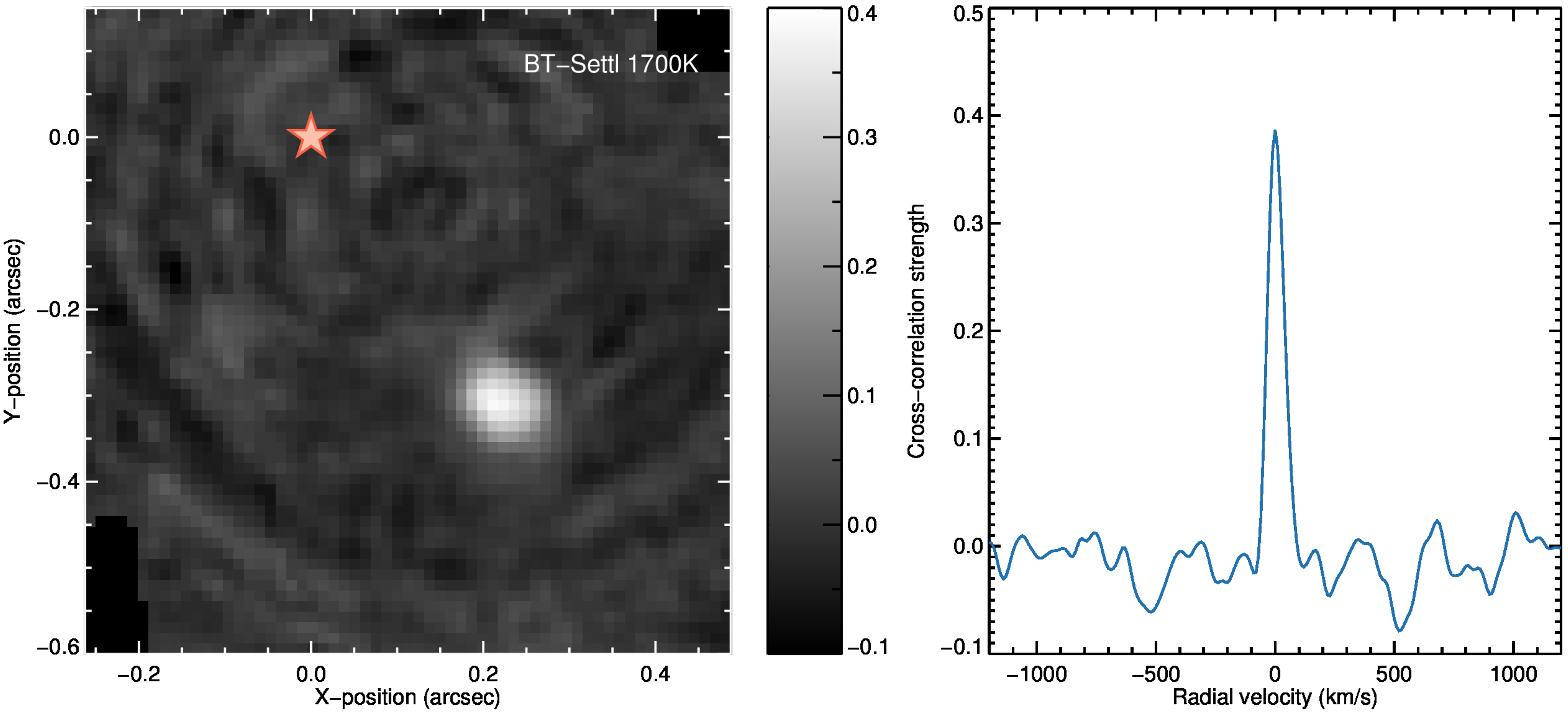}
   \caption{\textbf{Left panel:} Map of the cross-correlation function using the preferred BT-Settl model ($T_{\textrm{eff}}=1700\Kelvin$  and $\log (g)=3.5$).
 \textbf{Right panel:} The one-dimensional cross-correlation function at the location of the planet.}%
\label{fig:BTSettl_xcor}
\end{figure*}

The BT-Settl model that corresponds most closely to the planet parameters observed by \citet{Chilcote2017}, i.e. T$_{\rm{eff}}$=1700 K and $\log (g)=4.0$, is used to produce the cross-correlation map shown in the left panel of Fig. \ref{fig:BTSettl_xcor}. This model takes into account absorption from both \CO ~and \HTWOO ~and is therefore a more complete representation of the true planet spectrum than the individual molecule spectra. Indeed, the planet is detected at a significantly higher signal-to-noise ratio of \BTSIG. ~The K-band contrast ratio between $\beta$ Pic b and its host star is $\Delta K = 9.2\pm0.1$ \citep{Bonnefoy2011}. Our \BTSIG ~detection of the planet therefore corresponds to an achieved $3\sigma$ contrast of \CONTRAST, a factor \contrastratio ~deeper than the raw noise level measured in Section \ref{sec:observations} (see Fig. \ref{fig:contrast_curve}).

\subsection{Planet characterization using molecule mapping}
\begin{figure*}
   \centering
   \includegraphics[width=0.8\linewidth]{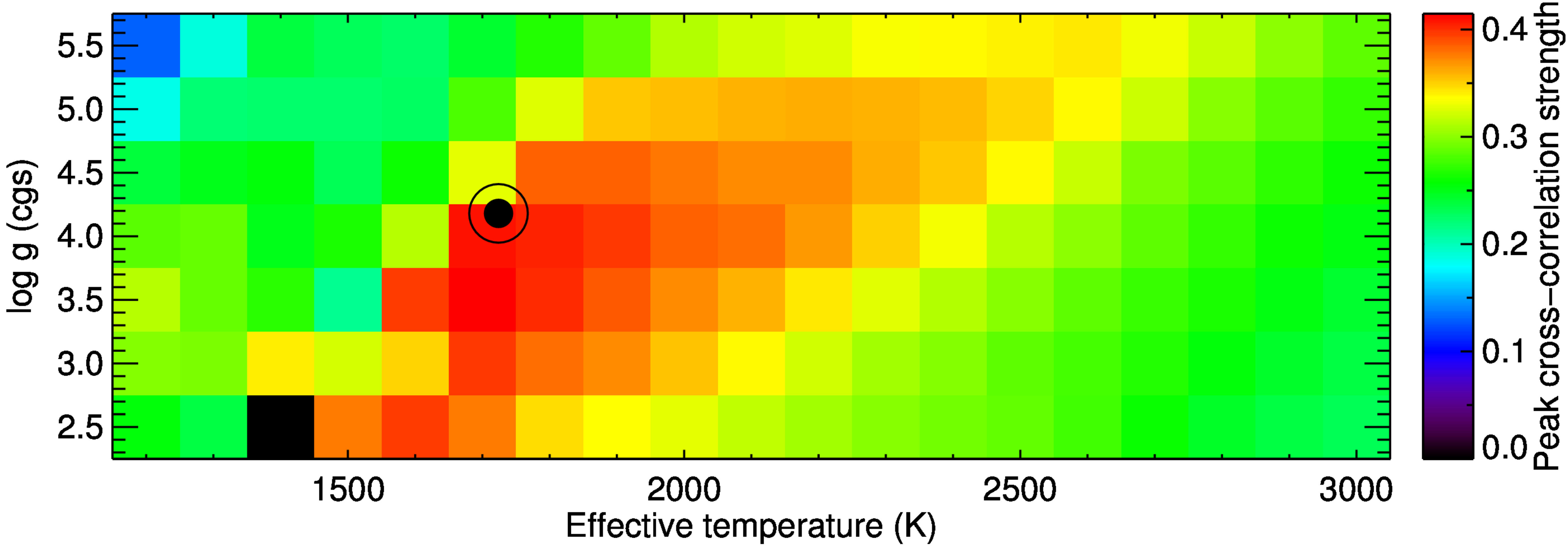}
   \caption{The planet cross-correlation peak as a function of BT-Settl models with varying $T_{\textrm{eff}}$  and $\log (g)$. The cross-correlation peak steeply decreases for temperatures below $\sim1600\Kelvin$ because the strength of the water absorption features at $>2.1\um$ become significantly weaker.
 All molecular bands diminish above temperatures of $\sim2000\Kelvin$, causing a slow decrease of the cross-correlation function at higher temperatures. The black circle corresponds to the values reported by \citet{Chilcote2017}.}
\label{fig:BTSettlgrid}
 \end{figure*}

Variations in global atmospheric parameters such as the effective temperature, surface gravity, metallicity and abundances ratios affect the relative strength of individual absorption lines. This may significantly influence the cross-correlation function, which means that the analysis is potentially sensitive to underlying model parameters. Fig. \ref{fig:BTSettlgrid} shows the peak value of the 1D cross-correlation function with BT-Settl models with varying $T_{\textrm{eff}}$  and $\log (g)$. The cross-correlation peak steeply decreases for temperatures below $\sim1600\Kelvin$ because the strength of the water absorption features at wavelengths larger than $2.1\um$ become significantly weaker at lower temperatures.

Above $\sim2000\Kelvin$ the cross-correlation peak is a shallower function of $T_\textrm{eff}$ as the \CO ~and \HTWOO ~absorption bands slowly diminish towards higher temperatures. Similarly, models with $\log (g)$ below 5.0 are favoured over models with higher surface gravities, due to the stronger water absorption bands relative to \CO ~for lower values of $\log (g)$. The highest cross-correlation peak is achieved for $T_{\textrm{eff}}=1700\Kelvin$  and $\log (g)=3.5$, with a signal-to-noise ratio of \BTSIG ~(see Fig. \ref{fig:BTSettl_xcor}). This result is in line with the values reported by \citet{Chilcote2017}, who obtain a consistent effective temperature and surface gravity by fitting hot-start evolutionary models \citep{Baraffe2003} to the bolometric luminosity of $\beta$ Pic b. Our analysis demonstrates that these parameters can also be obtained from the  medium resolution spectrum through cross-correlation. The somewhat higher surface gravity by \citet{Chilcote2017} is likely explained by the fact that these parameters are model dependent in both fitting approaches, and more work should be done to survey the possible inconsistencies across models, fitting methods and wavelength ranges.

More generally, we conclude that integral-field spectrographs can be used to characterize the fundamental parameters of directly imaged companions via spectral cross-correlation and that molecule mapping is therefore not limited to the detection of molecules in the atmospheres of these objects. Moreover, we predict that spectra covering a wider wavelength range that includes multiple water bands and bands of additional molecules will allow the effective temperature and surface gravity to be constrained significantly further using molecule mapping.

\subsection{Comparison with ADI-LOCI}
\begin{figure*}
   \centering
   \includegraphics[width=0.7\linewidth]{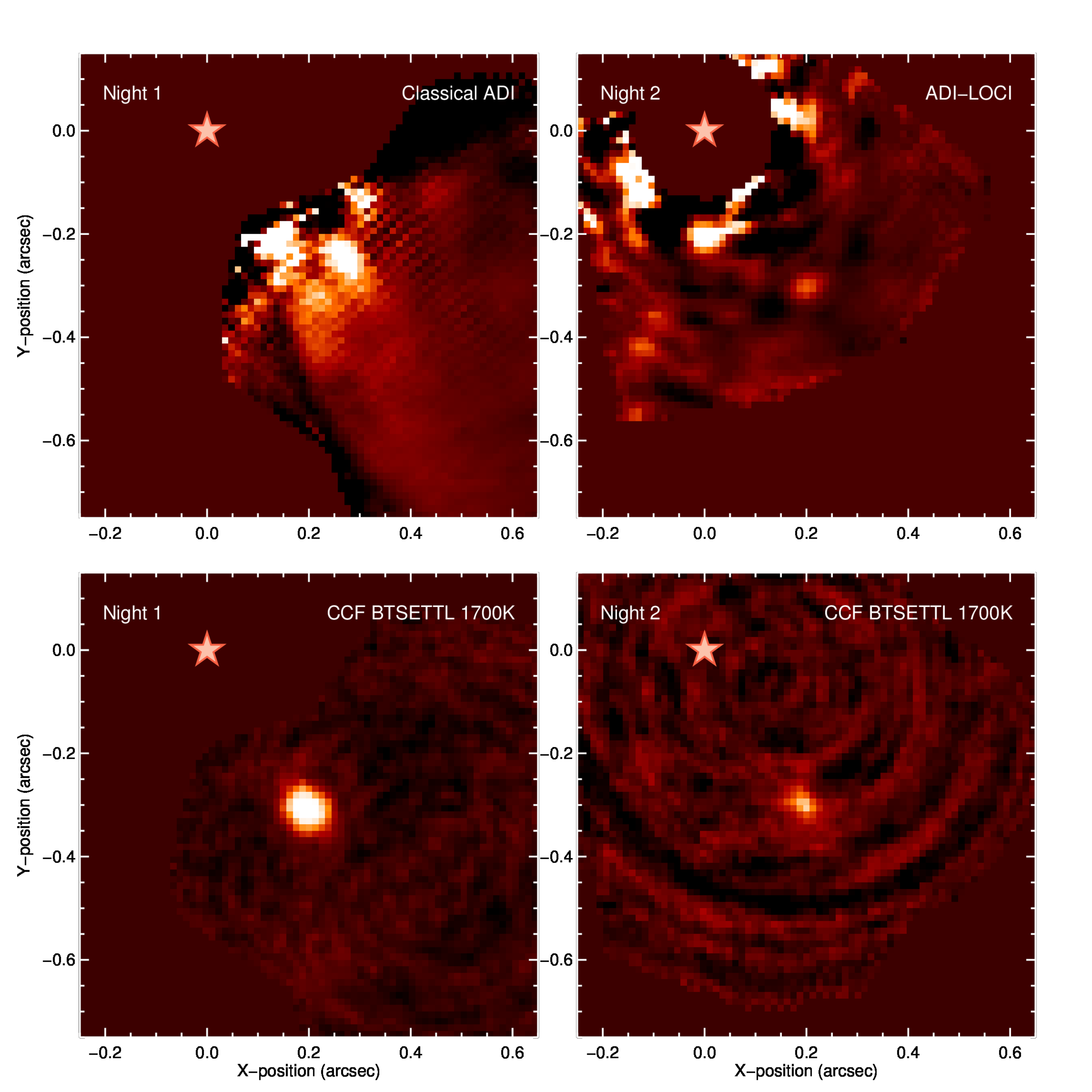}
   \caption{\textbf{Upper panels:} White-light images obtained after co-adding the data using classical ADI (left) and with the LOCI algorithm (right) for the first and second nights respectively. The planet is visible near [0.2",-0.3"], but is barely recovered in the second night, and not distinguishable from speckle noise in the first night. \textbf{Lower panels:} Cross-correlation maps using the preferred BT-Settl model ($T_{\textrm{eff}}=1700\Kelvin$  and $\log (g)=3.5$). The color scales linearly from black to white between -0.1 and +0.5 (left) and -0.05 to +0.25 (right).}
\label{fig:ADI_compare}
\end{figure*}

The upper panels of Fig. \ref{fig:ADI_compare} show the white-light images obtained using ADI, for which these observations were initially intended. These reveal that there is a strong difference in performance between the two observing nights: Because LOCI is hindered by field assymetry, the analysis of the data obtained during the first night is limited to classical ADI, which does not result in a detection of the planet. The planet is retrieved when using LOCI on the second night of data, but several other positive and negative features with similar amplitudes as that of the planet are present in the image, in particular close to the host star.

In contrast, the planet is detected most strongly in the molecule map of the observations taken in the first night, despite the field assymetry that hinders application of ADI. The observations taken in the second night have a much shorter exposure time than the first night, which was needed to place both the star and the planet in the field of view without saturating the detector\footnote{This effect is aggravated by the fact that the SINFONI detector shows strong persistence effects for count levels exceeding ~18.000 counts that dissipate over timescales up to 48 hours. The exposure time is therefore limited such that the nominal flux does not exceed 8.000 counts (see the SINFONI user manual for details).}. The standard day-time calibration plan of the SINFONI instrument includes the observation of three dark frames at each exposure time scheduled for use at night-time. This means that the dark current in this observing sequence is measured using only three dark frames with an exposure time of 2s each. We hypothesize that the observations of the second night are dominated by the combined read-noise and photon-noise in the master dark-frame and that this causes the strongly reduced sensitivity as compared to the first night. Future observations with SINFONI that target faint companions should therefore include a larger number of dark frames on top of those obtained in the standard calibration plan.

Nevertheless, our molecule mapping approach more effectively supresses the residual speckle pattern than ADI in both nights of data, leading to a stronger detection of $\beta$ Pic b and a higher sensitivity close to the star due to the absence of residual speckles.

\subsection{Molecule mapping from space with NIRSpec and MIRI}\label{sec:JWST}
Molecule mapping provides new possibilies for high-contrast imaging using medium-resolution IFSs, also using space-based observatories. Both NIRSpec and MIRI on the James Webb Space Telescope (JWST), will have integral-field capabilities. In IFS mode, NIRSpec\footnote{NIRSpec web documentation: https://jwst-docs.stsci.edu/display/JTI/Near+Infrared+Spectrograph\%2C+NIRSpec} covers wavelengths from 0.7 to $5.27\um$ at a spectral resolution of $\sim2.700$ with a spatial sampling of $0.1"\times0.1"$. MIRI has a similar IFS mode\footnote{MIRI web documentation: https://jwst-docs.stsci.edu/display/JTI/Mid-Infrared+Instrument\%2C+MIRI}, operating between $4.89\um$ to $28.45\um$ at a spectral resolution varying from 3320 at short wavelengths to 1460 at longer wavelengths. The waveband is divided into four channels with increasing slit-size, the bluest channel having a spatial resolution of $0.196"$, and the reddest a spatial resolution of $0.276"$. Although the spatial sampling of both NIRSpec and MIRI is significantly more coarse than that of SINFONI in 25 mas plate-scale mode, objects with a separation down to 0.1" can theoretically be resolved by the IFS of NIRSpec, and separations down to 0.2" can be resolved by MIRI at $5\um$.

\subsection{Molecule mapping with ERIS and HARMONI}\label{sec:HARMONI}
The ERIS integral-field spectrograph is expected to replace SINFONI at the Cassegrain focus of UT4 at the VLT in 2020 \citep{Amico2012,Kuntschner2014}. It will use an upgraded version of the existing IFS SPIFFY and a new infrared camera to achieve higher throughput and a spectral resolution of $R\sim8,000$ \citep{George2016}. In addition, it will employ a new wavefront module to enable significantly higher Strehl ratio's than SINFONI. These modifications will therefore increase the contrast and sensitivity to molecular signatures, making ERIS an important asset in the application of the molecule mapping technique until the arrival of the ELT.

The primary spectroscopic capability of the ELT in the near infra-red will be fulfilled by the HARMONI integral-field spectrograph. HARMONI will cover wavelengths between $0.47$ to $2.45\um$ at spectral resolutions ranging from $R\sim400$ to $\sim 20,000$ in individual Z, J, H and K bands. The highest possible spatial sampling is 4 mas, equivalent to the diffraction limit of the ELT \citep{Thatte2014}. 

To first order, the signal-to-noise achieved by cross-correlation scales with the signal-to-noise per wavelength element of the planet's absorption lines times the square root of the number of lines in the waveband \citep{Snellen2015}. For bright stars at near-infrared wavelengths, we can assume that the noise is dominated by the photon noise of the stellar PSF, the size of which which scales with the inverse of the mirror diameter $D$. The signal-to-noise of the planet spectral lines scales with the square root of the spectral resolving power, so the signal to noise of HARMONI compared to SINFONI can be approximated as:

$$\frac{S/N_{\textrm{HARMONI}}}{S/N_{\textrm{SINFONI}}} = \frac{D_{\textrm{ELT}}\sqrt{R_{\textrm{HARMONI}}}}{D_{\textrm{VLT}}\sqrt{R_{\textrm{SINFONI}}}} = 9.75$$

assuming the maximum spectral resolution of HARMONI, but otherwise equal wavelength coverage and instrument throughput. This indicates that HARMONI can achieve the same sensitivity as SINFONI in $\sim1\%$ of the exposure time, but for planets that are five times closer to their host star because the ELT mirror is five times bigger than that of the VLT.

With a spectral resolving power of 20,000 the systemic and orbital velocities of target planets can easily be resolved by the cross-correlation function if they exceed $15\kms$. In addition to discovering a companion and spectrally characterizing its atmosphere, the HARMONI instrument can therefore be used to establish co-movement with the host star and constrain the planet orbit via the instantaneous radial velocity. We conceive that this can be achieved in a single observation, making HARMONI especially suitable for detailed characterization of young giant exoplanets in short amounts of exposure time.

Mid-infrared coverage of the ELT is provided by METIS at high spectral resolution ($R\sim100,000$). Like HARMONI, METIS provides IFU capability for diffraction-limited imaging. It has already been recognized that this combination of high resolution spectroscopy with high contrast imaging gives METIS a unique capacity to characterize exoplanet atmospheres, using analysis strategies that are very similar to the one presented in this work \citep{Snellen2015}. Our work shows that the same strategy can be employed in the near-infrared at medium spectral resolution using HARMONI.

\section{Conclusion}\label{sec:conclusion}
This paper introduces the first application of the molecule mapping technique for detecting close-in substellar companions using AO-assisted medium-resolution integral field spectroscopy. With molecule mapping, integral-field spectra are cross-correlated with molecular template spectra to search for the spectral signatures of spatially resolved exoplanets that are embedded in the photon noise of the dominating star light. The cross-correlation co-adds the individual absorption lines of the planet spectrum at the spatial location of the planet, but not (residual) telluric and stellar features. This acts to suppress the quasi-static speckle pattern that is a limiting factor in standard direct-imaging analyses.

We applied molecule mapping to 2.5 hours of archival SINFONI observations of the $\beta$ Pictoris system. By cross-correlating the integal-field data cubes with templates of \CO, \HTWOO, \CHFOUR ~and \NHTHREE, ~we obtained signal-to-noise ratios for \CO ~and \HTWOO ~in the atmosphere of the planet of \COSIG ~and \WATERSIG ~respectively. We also cross-correlated with a grid of BT-Settl models, varying the effective temperature between $1200 \Kelvin - 3000 \Kelvin$, and the surface gravity between $\log (g) = 2.5 - 5.5$. We found that the cross-correlation peaks for $T_\textrm{eff} = 1700 \Kelvin$ and $\log (g) = 3.5$, which shows that molecule mapping can also constrain the fundamental parameters of young gas giant planets. With these model parameters, the planet was detected at a signal-to-noise ratio of \BTSIG, ~corresponding to a $3\sigma$ contrast of \CONTRAST. ~We also analysed the same data using ADI, resulting in only a marginal detection and a strong difference in sensitivity in the two individual nights, due to the different observing strategies used and the associated calibrations. However in both nights of data molecule mapping outperformed ADI, demonstrating that it can significantly enhance the sensitivity of IFS observations at medium spectral resolution.

Our successful application of molecule mapping on existing SINFONI data feeds expectations for the potential of upcoming medium-resolution integral-field instruments on the JWST and the ELT. We briefly outlined the specifications of MIRI and NIRSpec (JWST), ERIS (VLT) and HARMONI (ELT), and anticipate that these instruments are well suited for such cross-correlation based analyses.

\begin{acknowledgements}
       This work is part of the research programme VICI 639.043.107 funded by the Dutch Organisation for Scientific Research (NWO). I. Snellen acknowledges funding from the European Research Council (ERC) under the European Union’s Horizon 2020 research and innovation programme under grant agreement No 694513. This work is based on observations collected at the European Organisation for Astronomical Research in the Southern Hemisphere under ESO programme 093.C-0626(A). Finally, G. Chauvin, M. Bonnefoy and A.-M. Lagrange acknowledge support from the French National Research Agency (ANR) through project grant ANR10-BLANC0504-01 and the Programmes Nationaux de Planétologie et de Physique Stellaire (PNP \& PNPS), in France.
\end{acknowledgements}

\bibliographystyle{aa} 
\bibliography{bib} 

\end{document}